\documentclass{article}%
\usepackage{amsmath}
\usepackage{amsfonts}
\usepackage{amssymb}
\usepackage{graphicx}%
\providecommand{\U}[1]{\protect\rule{.1in}{.1in}}
\begin{document}

\title{Dark matter and Wigner's third positive energy representation class}
\author{Bert Schroer\\permanent address: Institut f\"{u}r Theoretische Physik\\FU-Berlin, Arnimallee 14, 14195 Berlin, Germany\\present address: CBPF, Rua Dr. Xavier Sigaud 150, \\22290-180 Rio de Janeiro, Brazil}
\date{August 30, 2013}
\maketitle

\begin{abstract}
Positive energy representations of the Poincar\'{e} group are naturally
subdivided into three classes according to their mass and spin content: m%
%TCIMACRO{\TEXTsymbol{>}}%
%BeginExpansion
$>$%
%EndExpansion
0, m=0 finite helicity and m=0 infinite helicity. For a long time the the
quantum field theory of the third class remained a mystery before it became
clear that one is confronted with a new kind of "stuff" with very different
properties from matter as we know it: unlike normal matter it cannot be
localized in compact spacetime regions and its generating quantum fields are
semi-infinite spacelike strings. In this note we present arguments that such
noncompact stuff is inert apart from gravitational coupling which makes it a
perfect candidate for dark matter.

\end{abstract}

\section{String-localization and Wigner's infinite spin representation}

Wigner's famous 1939 theory of unitary representations of the Poincar\'{e}
group $\mathcal{P}$ was the first systematic and successful attempt to
classify relativistic particles without relying on a Lagrangian quantization
parallelism to classical field theory \cite{Wig}. As we know nowadays,
Wigner's the massive and the massless finite helicity class of positive energy
representations of $\mathcal{P}$ cover all known particles and their
descriptions in terms of free fields. Coupling these covariant pointlike free
fields to form scalar interaction densities is the starting point of
perturbation theory on which our present understanding of interacting matter
is based.

Whereas the massive representation class $(m>0,s=\frac{n}{2})$, covers all
known massive particles, the massless representations split into two classes
which belong to very different unitary representations of the little group.
The latter is the invariance group of a lightlike vector i.e. the
three-parametric Euclidean group $E(2)$ in two dimensions. Its degenerate
representations are the ($m=0,\pm\left\vert h\right\vert $) two-component
massless helicity representations, whereas the faithful $E(2)$ representations
are infinite dimensional and define the third class of positive energy
representations; they were referred to by Wigner as "infinite spin"
representations \footnote{In the more recent literature they are sometimes
(more appropriately) called "continuous spin" representations.}.

The localization properties of this third class turned out to be incompatible
with pointlocal fields \cite{Y}. There are numerous failed attempts which
tried to enforce pointlike localization; some of them are mentioned in the
first volume of Weinberg's book. Weinberg himself concluded that "nature does
not use it". In view of the fact that in those days dark matter was not yet an
issue of high energy physics and that all known particles had been identified
within Wigner's finite spin/helicity classes, this was a factually correct statement.

All known and conjectured zero mass particles (photons, gravitond, possibly
massless neutrinos) belong to the finite helicity class. In formulating
interactions of vectormesons in a Hilbert space one has to avoid pointlike
vectorpotentials (which inevitably leads to gauge theory in indefinite metric
Krein spaces and the presence of ghost operators). The Hilbert space
positivity requires to use covariant stringlocal vectormesons \cite{Hilbert}
but there are still pointlike generated local observables (corresponding to
gauge-invariant pointlike fields). The QFT of the third Wigner class is
totally different in that \textit{all} (including composite) fields are
stringlocal i.e. the localization of the Wigner "stuff" is wholly
\textit{noncompact}. We avoid on purpose the terminology "nonlocal" since this
stuff admits causal localization in arbitrary narrow spacelike cones; it only
cannot be causally localized in compact spacetime regions. To be more precise,
in the QFT of ordinary matter there exist useful global observables as
conserved charges. But they are always limits of sequences of compact
localized operators, whereas the Wigner stuff is "irreducibly" noncompact and
defines a completely new type of QFT.

It turns out that the new Hilbert space description of Yang-Mills couplings
contains objects whose properties stand in an interesting and fruitful
contrast to the Wigner stuff, namely confined gluons and quarks. The
interaction of massive gluons and quarks leads to a renormalizable
interactions e.g. to massive physical stringlocal gluons which are quite
different from gauge-variant pointlike gluon fields. Analogies to the
Yennie-Frautschi-Suura \cite{YFS} treatment of logarihmic on-shell infrared
divergencies in QED and their re-summation techniques for leading logarithic
divergences in $m\rightarrow0$ suggest that all correlation functions
containing in addition to pointlike composites stringlocal vectormesons or
quarks\footnote{Apart from $q-\bar{q}~$configurations in which the large
distance parts of the two strings cancel and the remaining finite
string-bridge is parallel to the spacelike separation of the pair.} vanish; in
fact this property should be taken as the definition of confinement. This
property scotches the undesired occurance of an acausal process in which
colliding compact matter creates noncompact stuff. Whereas gluons and quarks
cannot come out of such a process (and therefore do not show up in the
energy-momentum balance), the noncompact Wigner stuff, once in this world,
cannot change into normal matter, which accounts for its inert behavior.

It is not that the problem of the QFT of Wigners infinite spin representation
class was ignored after Wigner's unsucessful attempts. There were several
later equally unsuccessful attempts to press the covariant content of this
representation into the form of a pointlike field. The first hint into the
right direction was Yngvason's theorem establishing that this representation
is incompatible with pointlike localization \cite{Y} The group theoretical
covariantization method for Wigner's unitary representation theory as used in
Weinberg's book did not resolve the problem, and Weinberg's dictum was that
nature does not use this representation since matter as we know it can be
taken care off in terms of massive or massless finite helicity
representations. This was factually correct since the concept of dark matter
did not enter the consciousness of particle theorists at that time.

With the help of the more recent intrinsic "modular localization" concept this
problem was finally solved in two steps. First the application of this idea to
the construction of modular localized subspaces of Wigner's representation
space revealed that \textit{all positive energy representations can be
localized in arbitrary narrow spacelike cones} (whose cores are semi-infinite
spacelike strings) \cite{BGL}. Apart from the third representation class, this
noncompact localization can be sharpened to compact (double cone) regions with
pointlike generating wave functions; this is however not possible for the
third class since the compact localized subspaces turn out to be trivial. The
core of arbitrary small double cones is a point and that of spacelike cones a
string. Since the generating covariant wave functions or quantum fields in the
first two cases are known to be pointlike, one expected the covariant third
class fields to be stringlocal. The associated stringlocal covariant fields
were explicitly constructed in \cite{MSY}. This did not stop the futile
attempt to relate the infinite spin Wigner representation with pointlike
fields \cite{Schu} more than 40 years after a No Go theorem to this effect had
been established \cite{Y} and 7 years after the appropriate stringlocal fields
were constructed \cite{MSY}; an unmistakable sign of increasing fragmation of
individual knowledge about particle physics in times of globalization.

The absence of compact localizability for the somewhat mysterious third
positive energy representations class has radical physical consequences. Such
inert stuff cannot be registered in a counter, neither can it be generated
from a collision of ordinary matter. Its main property is its reactive
inertness which manifests itself in the absence of almost all properties of
ordinary matter, except its coupling to gravity as a consequence of the
positive energy property. The arena of action of noncompact matter are
galaxies and not earthly laboratories. The astrophysical arguments against
identifying massless ordinary matter are not applicable to the noncompact
third class stuff. Unlike normal localizable massless matter it cannot escape
from galaxies, rather it pervades them and extends into the empty space. Its
main and, as will be argued in this paper, only manifestation is the change
its presence causes in the galactic gravitational balance.

The identification of the third class Wigner stuff with dark matter has two
virtues as compared to the other proposals. It is not an ad hoc invention for
explaining dark matter; Wigner's classification is as old as Zwicky's
astrophysical observations which led to the dark matter proposal. Furthermore
it fulfills the requirement of the possibility of its falsification since any
convincing identification of a counter registration event with the
astrophysical dark matter disproves the present proposal; the third class
Wigner stuff turns out to be inert par excellence, apart from its
gravitational coupling.

A lot about physical properties of noncompact matter can be learned by
confronting its stringlocal quantum fields with new insights about higher spin
ordinary matter \cite{stringlocal} \cite{Hilbert}. By this we specifically
mean the recent discovery of the necessity of using \textit{stringlocal fields
in order to maintain the Hilbert space positivity for renormalizable higher
spin }$s\geq1~$\textit{\ interactions}\footnote{The standard setting for
renormalizable $s=1$ interactions is the pointlike Becchi-Rouet-Stora-Tuytin
(BRST) formulation of operator gauge theory which replaces Hilbert space by an
indefinite metric Krein space.}. The use of $s=1~$stringlocal fields in
Hilbert space (instead of pointlike fields in Krein space) promises to lead to
significant progress in the understanding of infrared problems (as e.g.
confinement) which arise in the massless limit of interacting physical
(Hilbert space) vectormesons. In fact the perception about the relevance of
the use of stringlocal vectormesons for maintaining a Hilbert space
description arose in the aftermath of the discovery of the field theoretic
description of the infinite spin Wigner stuff and the question arises to what
extend one may be able to use this connection in the opposite direction.

Massless $s=1$ interactions are best understood in terms of massless limits of
correlation functions of interacting massive stringlocal vectormesons. This is
because QFTs with a mass gap have a rather simple relation between fields and
particles which manifests itself in the validity of time dependent (LSZ)
scattering theory and the ensuing Wigner-Fock structure of the Hilbert space.
These properties become blurred in the massless limit and lead to perturbative
infrared divergencies whose physical consequences can be investigated by YFS
resummation techniques of leading infrared divergencies \cite{YFS}. The analog
for the Wigner stuff would be to represent its stringlike fields as a massless
limit of a massive high spin representation in which the vanishing of the
decreasing mass is coupled to a growing spin. Unfortunately it is presently
not known how to do that or even if this is possible.        

Falling short of such an ambitious goal, the main point of the present work is
to use what has been learned from the Hilbert space reformulation of the BRST
gauge theory for a better understanding of possible physical manifestations
(or rather their absence, apart from the gravitational coupling) of the Wigner
stuff as a consequence of its known very strong semi-infinite
string-localization. A placative formulation of the result would consist in
viewing darkness and confinement as the two opposite flanks of normal matter:
confinement is connected with a property of particular interacting stringlocal
fields which diappears in the $m\rightarrow0$ limit of stringlocal massive
vectormesons, whereas dark matter, once in this world, cannot disappear by
being converted into normal matter\footnote{Whereas Inert matter carries
energy-momentum, confined fields/particles i.e. fields which vanish in the
massless limit, do not contribute to the energy-momentum balance. }. The
foundational reasons, namely the havoc with the causality principle which a
change of compact spacetime localized matter into irreducible noncompact stuff
would cause, are shared but the physical consequences are very different.

Some details on the connection between Wigner's positive energy
representations with localization can be found in the next two sections. The
last section comments on the use of the inert Wigner stuff as a possible
explanation behind the observed astrophysical darkness.

\section{Point- and string-like generating fields of positive energy
representations}

Recent progress on foundational localization problems revealed that the
generating fields of the infinite spin Wigner stuff representation class is
stringlocal. We will only present the results and refer the reader for their
derivation to \cite{MSY} \cite{Pla}. For the selfconjugate bosonic case (to
which we limit our exemplary presentation) they are of the form%

\begin{align}
&  \Psi(x,e)=\frac{1}{\left(  2\pi\right)  ^{\frac{3}{2}}}\int\{e^{ipx}%
u^{\alpha}(p,e)\cdot a^{\ast}(p)+e^{-ipx}\overline{u^{\alpha}(p,e)}\cdot
a(p)+\}\frac{d^{3}p}{2\omega}~~\label{cov1}\\
&  u\cdot\bar{u}:=\int d^{2}k\delta(k^{2}-\kappa^{2})u(k)\bar{u}%
(k),~~\ u^{\alpha}(p,e)(k)=e^{-i\pi\alpha/2}\int d^{2}ze^{ikz}(B_{p}%
\xi(z)\cdot e)^{\alpha},\nonumber\\
~  &  \xi(z)=(\frac{\left\vert z\right\vert ^{2}+1}{2},z_{1},-z_{2}%
,\frac{\left\vert z\right\vert ^{2}-1}{2}),~~e^{\mu}e_{\mu}=-1\nonumber\\
&  (D^{\kappa}(c,R(\theta))\varphi)(k)=e^{ic\cdot k}\varphi(R^{-1}%
(\theta)k),~\mathfrak{h=~}L^{2}(\mathbb{R}^{2},\delta(k^{2}-\kappa^{2}%
)d^{2}k)\nonumber
\end{align}
Here the intertwiners $u^{\alpha}(p,e)(k)~$are $p,e$ -dependent functions on
the two-dimensional $k$-space on which the 2-dim Euclidean group $E(2)$ acts
($c~$translation, $\theta$ rotation) and in this way defines a representation
on a Hilbert space $\mathfrak{h}$. The Pauli-Lubanski invariant $\kappa\;$is a
continuous parameter ("the continuous spin/helicity") which characterizes the
$E(2)$ representation and defines a Casimir invariant of the positive energy
representation of $\mathcal{P}$ associated to the stuff. The $u$ are called
"intertwiners" because they convert the unitary Wigner representation acting
as the adjoint representation on $a^{\#}(p)(k)~$into the covariant
transformation of the $\Psi(x,e).$ They also lead to stringlike causality i.e.
the vanishing of the commutator for spacelike separations of the two
strings\footnote{We only need the simplest realization of Wigner strings
(bosonic, self-conjugate, without additional spinorial indices). Covariant
strings are necessarily straight.} in relative spacelike positions
\begin{align}
&  U(\Lambda,a)\Psi(x,e)U(\Lambda,a)^{\ast}=\Psi(\Lambda x+a,\Lambda
e)\label{cov}\\
&  \left[  \Psi(x,e),\Psi(x^{\prime},e^{\prime})\right]  =0,~for~x+\mathbb{R}%
_{+}e><x^{\prime}+\mathbb{R}_{+}e^{\prime}\nonumber
\end{align}
The derivation of the formula for the covariant intertwiners uses ideas from
modular localization \cite{MSY} in addition to the group theoretic properties
which were already used for the construction of the quantum fields for the
massive and finite helicity massless class by Weinberg \cite{Wein}.

The transcendental $u$-intertwiners lead to rather involved two-point
functions and propagators%
\begin{align}
&  \left\langle \Psi(x,e)\Psi(x^{\prime},e^{\prime})\right\rangle =\frac
{1}{\left(  2\pi\right)  ^{3}}\int e^{-ip(x-x^{\prime})}M(p;e,e^{\prime
})~~\label{prop}\\
&  M(p;e,e^{\prime})=\int u(p,e)(k)\overline{u(p,e^{\prime})}(k)\delta
(k^{2}-\kappa^{2})d^{2}k\frac{d^{3}p}{2p_{0}}\\
&  \left\langle \Psi(x,e)\Psi(x^{\prime},e^{\prime})\right\rangle
\rightarrow\left\langle T\Psi(x,e)\Psi(x^{\prime},e^{\prime})\right\rangle
~~~by~~\frac{d^{3}p}{2\left\vert \vec{p}\right\vert }\rightarrow\frac{1}{2\pi
}\frac{1}{p^{2}-i\varepsilon}d^{4}p\nonumber
\end{align}
where the third line denotes the transition from the two-pointfunction to the
propagator by changing the p-integration.

For later use we also present the corresponding representations for pointlocal
fields from the massive and finite helicity Wigner class ($b$ refers to antiparticles)%

\begin{equation}
\psi^{A,\dot{B}}(x)=\frac{1}{\left(  2\pi\right)  ^{3/2}}\int(e^{ipx}%
u^{A,\dot{B}}(p)\cdot a^{\ast}(p)+e^{-ipx}v^{A,\dot{B}}(p)\cdot b(p))\frac
{d^{3}p}{2p_{0}} \label{Wein}%
\end{equation}
The intertwiners $u(p)$ for $m>0$ are rectangular $(2A+1)(2B+1)><(2s+1)$
matrices which intertwine between the unitary $(2s+1)$-component Wigner
representation and the covariant ($A.\dot{B}$) spinorial representation; the
$a,b$ refer to particle and antiparticle creation/annihilation operators. For
the $m=0$ representations the formula is the same, except that dot stands for
the inner product in a two-dimensional space (the space of the two helicities
$\pm\left\vert h\right\vert $). Another difference between the massive and the
massless case is the range of possible spinorial indices; for a given physical
spin $s~$the range of spinorial (half)integer spinorial representation indices
of the homogeneous Lorentz group is restricted by%
\begin{align}
\left\vert A-\dot{B}\right\vert  &  \leq s\leq A+\dot{B},~~m>0\\
\left\vert A-\dot{B}\right\vert  &  =\left\vert h\right\vert ,~~~m=0
\label{cov2}%
\end{align}
the second formula shows that the the vector representation $A=1/2=B$ does not
occur for $m=0$~i.e. \textit{pointlike massless covariant vectorpotential are
not consistent with the Hilbert space positivity of quantum theory }(the
mentioned clash of massless $s\geq1~$pointlike tensorpotentials with Hilbert
space positivity).

As mentioned, the discovery of the stringlocal noncompact third class Wigner
stuff was the beginning of a systematic study of \textit{stringlocal} fields
for the two pointlike generated finite spin/helicity representation classes
associated with compact localizable ordinary matter. The absence of pointlike
massless vectorpotentials (more generally $s\geq1$ tensor potentials) led to
the definition of covariant stringlocal vectorpotentials with the covariance
and locality property (\ref{cov})
\begin{align}
&  A_{\mu}(x.e):=\int_{0}^{\infty}e^{\nu}F_{\mu\nu}(x+se)ds,~~e^{\mu}e_{\mu
}=-1\label{cov3}\\
&  \left\langle A_{\mu}(x,e)A_{\nu}(x^{\prime},e^{\prime})\right\rangle
=\frac{1}{(2\pi)^{3/2}}\int e^{-ip(x-x^{\prime})}M_{\mu\nu}(p;e.e^{\prime
})\frac{d^{3}p}{2p_{0}}\nonumber\\
&  M_{\mu\nu}(p;e.e^{\prime})=-g_{\mu\nu}-\frac{p_{\mu}p_{\nu}e\cdot
e^{\prime}}{(p\cdot e-i\varepsilon)(p\cdot e^{\prime}+i\varepsilon)}%
+\frac{p_{\mu}e_{\nu}}{p\cdot e-i\varepsilon}+\frac{p_{\nu}e_{\mu}^{\prime}%
}{p\cdot e^{\prime}+i\varepsilon}\nonumber
\end{align}
where the $\varepsilon$-prescrition defines the two-point distributions as
boundary values of analytic functions. Different from the pointlike Proca
potential, these stringlocal fields permit a massless limit \cite{stringlocal}
\cite{Hilbert}.

There is no problem with the existence of pointlike $s\geq1$ higher spin
potentials, apart from the fact that their short distance dimension increases
with spin $d_{sd}=s+1.~$Already for $s=1$ the $d_{sd}=2~$pointlike$~$Proca
vectorpotential $A_{\mu}^{P}(x)$ is too singular in order to permit a
renormalizable interaction density within the power-counting limit
$d_{sd}^{int}\leq4.$ But the massive stringlocal vectorpotential constructed
according to (\ref{cov3}) has $d_{sd}=1,$ in fact \textit{all} $d_{sd}=s+1$
pointlike tensor potentials have stringlocal siblings of $d_{sd}=1.~$The
field-fluctuations in the directional spacetime variable $e$ has led to a
reduction of the strength of $x$-fluctuations from $s+1$ to $1.$

For $s=1$ and $m>0$ there exists also a $d=1~$stringlike scalar%
\begin{align}
&  \phi(x,e)=\int_{0}^{\infty}e^{\mu}A_{\mu}^{P}(x+se)ds\label{cov4}\\
F_{\mu\nu}  &  :=\partial_{\mu}A_{\nu}^{P}-\partial_{\nu}A_{\mu}^{P}%
,~~~A_{\mu}(x,e):=\int_{0}^{\infty}e^{\nu}F_{\mu\nu}(x+se)ds
\end{align}
The second line presents the stringlocal field in terms of its pointlike Proca
sibling so that its two.pointfunction (\ref{cov3}) follows from that of the
Proca field; all other two-pointfunctions, including the mixed $A$-$\phi$
ones, can be derived via (\ref{cov4}) from that of the Proca field. In fact
the 3 fieds obey the linear relation%
\begin{equation}
A_{\mu}(x,e)=A_{\mu}^{p}(x)+\partial_{\mu}\phi(x,e) \label{rel}%
\end{equation}
in which the explicit dependence of line integrals has been absorbed into the
defintion of stringlocal fields. The relation of $d_{sd}=1$ stringlocal
tensorpotentials of spin $s$ to their pointlike siblings requires the presence
of $s$ lower spin $\phi^{\prime}s$.

It turns out that the stringlocal scalar $\phi$ plays an important role in the
formulation of stringlike interactions; it "escorts" the $A_{\mu}%
(x,e)~$potential and enters explicitly the interaction densities, although it
does not add new degrees of freedom\footnote{It plays an important role in the
coupling of a massive vectormeson to a Hermitian $H$-field which turns out to
be the correct formulation (no symmetry-breaking, no spontaneous creation of
vectormeson masses) of the Higgs model \cite{stringlocal}.}. All three fields
are linear combinations of the three $s=1$ Wigner creation/annihilation
operators $a^{\#}(p,s_{3})$ with different $u$-intertwiners; in fact
(\ref{rel}) is nothing else than a linear relation between three intertwiners.
Pointlike intertwiners are matrices with \textit{polynomial} entries in $p$ of
degree $d_{sd}=s$ whereas the $p$-$e$ dependence of stringlocal tensor
potentials and their lower spin $\phi^{\prime}s~$is \textit{rational} in
$p,e$. Such stringlocal free fields are "reducible" in the sense that they can
be written as semi-infinite line integrals over pointlike observables. This is
to be compared to the transcendental $p$-$e$ dependence of the third class
intertwiners (\ref{cov1}) which represents \textit{irreducible} strings whose
smearing in $x,e~$results in noncompact (referring to spacetime localization)
stuff. Nonlocal objects as e.g. conserved global charges also appear in normal
compact localizable QFTs, but they can always be described as global limits of
local operators, except for the Wigner stuff where such a local approximation
is not possible.

This picture about free fields and their associated particles changes
drastically in the presence of interactions. Whereas interactions involving
massive $d_{sd}=s+1$ pointlike fields are nonrenormalizable, the
transformation of such interactions into their stringlike counterpart permits
to construct renormalizable interactions for any spin. The stringlike
renormalizable formulation shows that \textit{behind the pointlike failure of
renormalizability is a weakening of localization in the sense of nonexistence
of pointlike Wightman fields whose role is taken over by renormalizable
stringlocal fields}; in short, the breakdown of pointlike renormalizability is
caused by the weakening of localizability from compact to noncompact
localization. This is interesting because it shows that the weakening of
localizability is interwoven with a radical worsening of pointlike short
distance behavior which manifests itself in a breakdown of renormalizability.
Wightman localizability (operator-valued Schwartz distributions) amounts to
polynomial boundedness in momentum space; this can only be restored by the
reformulation in terms of stringlike fields.

It turns out that behind the point- versus string-like localization is the
powerful Hilbert space positivity: \textit{the stringlike localization is the
tightest localization which is consistent with Hilbert space positivity} i.e.
for generating the net of localized algebras which is the algebraic
description of QFT \cite{Haag} one does not need generating fields which are
localized e.g. on spacelike hypersurfaces. Here pointlike is viewed as a
special case of stringlike (i.e. pointlike $\simeq$independence on $e$).
Renormalization theory shows that pointlike renormalizability and the absence
of massless infrared divergencies is limited to $s<1,$ whereas
renormalizability if interactions involving $s\geq1~$fields requires
string-localization and may lead (depending on the interaction) to infrared
divergencies in the massless limit \cite{stringlocal}. In other words the
origin of infrared divergencies (which only start to appear for $s\geq1$
interactiong fields\footnote{Couplings of $s<1$ fields with interactions
within the power-counting limitation have well-defined infrared-divergence
free massless limits.}) is the long range interaction caused by massless
stringlike fields. Pointlike interactions for $s<1~$are compatible with the
Hillbert space and do not cause infrared divergencies.

At this point the attentive reader may want to know how quantum gauge theory
fits into this new setting. Quantum gauge theory abandons the Hilbert space
positivity and keeps instead the pointlike formalism. Whereas the classical
gauge formalism fits well into classical field theory, its quantum counterpart
violates the Hilbert space positivity (which is the Holy Grail of quantum
theory). Though this property is later recovered for the gauge-invariant
fields whose application to the vacuum create a smaller Hilbert space (the
vacuum sector), the formalism does not provide physical operators whose
application to the vacuum describe charged states; in fact the unphysical
nature of gauge-dependent allegedly charge-carrying matter fields is evident
from the observation that the Maxwell charge of the associated states (which
they create from the vacuum) vanishes \cite{Froe}. The 70 year use of quantum
gauge theory, which entered QFT through Lagrangian quantization, and the
discovery of successful recipes to navigate around these shortcomings (viz.
the photon-inclusive cross sections in QED, the prescriptions of physical
hadrons in terms of composites of gauge-dependent quarks) led to the loss of
awareness about its limitations. From a mathematical viewpoint perturbation
theory in an indefinite metric (Krein space) setting is pure combinatorics
outside the range of functional analytic or operator algebraic control
(violation of the Cauchy-Schwarz inequality,,..).

The new stringlocal setting maintains the powerful positivity restriction
coming with the Hilbert space and with it the applicability of operator
algebraic methods at a seemingly small price of weakening of localization from
point- to string-like. It opens the path to the construction of stringlocal
physical matter- and Yang-Mills fields and establishes the conceptual
prerequisites for studying the remaining important open problems of QFT
including the unsolved problem of confinement. In the present work QFT is
always meant in a Hilbert space setting unless the Krein space gauge theory
setting is explicitly mentioned.

The reason for recalling recent results about the use of stringlocal fields
(which are presently revolutionizing our ideas about localization properties
and their physical manifestations of normal matter \cite{stringlocal}
\cite{Hilbert}), is that the best characterization of the infinite spin Wigner
stuff is in terms of the \textit{absence} of most properties of matter as we
know it. This problem will be taken up in the next section.

\section{Wigner's infinite spin stuff and matter as we think we know it}

Since causal localization is the foundational property of QFT, the main task
of particle physics is to explain the wealth of observed properties as
different manifestations of this unifying principle in the context of
different models of QFT. It requires in particular to understand in which
sense confinement in QCD and spacetime properties of infraparticles in QED are
related to localization. The Hilbert space setting suggest a clear picture
behind such infrared problems.

Its starting point is the mass gap property which secures the Wigner-Fock
particle structure of the Hilbert space. The resulting field-particle relation
is described by a structural theorem which states that in a theory with local
observables (which define the vacuum sector) the superselection-charge
carrying in/out scattering states can be described in terms of time-dependent
LSZ scattering theory applied to operators localized in arbitrary narrow
spacelike cones (whose cores are strings) \cite{Bu-Fr}. In this case the new
perturbation theory based on stringlocal fields (SLF) permits the construction
of \textit{singular} polynomially unbounded pointlike fields whose direct
perturbative use would have led to nonrenormalizability \cite{stringlocal}.

This changes abruptly in the limit of massless vectormesons. In case of QED
the use of the Gauss theorem (appropriately adapted to QFT \cite{Bu}) shows
that the asymptotic direction of the spacelike cone-localization of
charge-carrying operators can not be changed by unitary operators; in other
words the directions $e$ of the generating semi-infinite stringlocal fields
are "rigid". In particular the Lorentz covariance outside the vacuum sector is
spontaneously broken \cite{Froe} and the culprits are soft photon clouds which
hover along the spacelike semi-line $x+\mathbb{R}_{+}e.$

This accounts for a change in the field-particle relation; in particular the
mass-shell of the charged particle which in theories with mass-gap leads to a
$\theta(p_{0})\delta(p^{2}-m^{2})~$contribution (or a $(p^{2}+m^{2})^{-1}$
pole contribution in the time-ordered functions) "dissolves" into the
continuum in form of a milder cut singularity with threshold starting on the
mass-shell. In this case a spacetime dependent collision theory of
infraparticles which generalizes the LSZ scattering theory\footnote{The
application of LSZ would lead to vanishing large time asymptotic limits since
the weaker threshold singularity cannot compensate the dissipation of wave
packets.} does not yet exist and one has to take recourse to the well-known
successful \textit{prescription} in terms of photon-inclusive cross-sections
\cite{YFS}. The perturbative manifestations are the well-known on-shell
logarithmic infrared divergencies whose resummation in leading orders lead to
power behavior in the infrared cutoffs. The removal of the infrared
regularization implies the vanishing of scattering amplitudes which can be
avoided by passing to the photon-inclusive cross-sections before removing the
infrared cutoff \cite{YFS}. The new stringlocal setting promises to lead to a
spacetime understanding of these prescriptions in which the ad hoc
noncovariant infrared regulator is replaced by the more natural mass of the
stringlocal vectormeson while the unphysical (gauge-dependent) pointlike
matter fields pass to stringlocal charge-carrying fields in Hilbert space.

As mentioned above the situation changes in the limit of massless gluons.
There remains a significant difference between QED and QCD; in QED the
vectorpotential can be written as a semi-infinite line integral over a
pointlocal observable (the field strength), whereas stringlocal interacting
gluon fields cannot be written in this way. In the latter case the strings are
"irreducible". In the massless limit the singular pointlike siblings of the
stringlocal fields disappear and the interacting massless gluon matter becames
inherently noncompact, whereas certain $e$-independent composites generate
(corresponding to the gauge-invariant observables of gauge theory) the compact
localizable part of Y-M matter. If such intrinsic noncompact matter could
emerge from a collision of compact matter one would have serious problems with
causality. This suggests to define gluon/quark confinement of particle theory
as the \textit{vanishing of correlation functions which contain} besides
composite pointlike observable fields also \textit{stringlocal gluon and quark
fields}\footnote{The only exception are $q-\bar{q}$ pairs in which the
$e$-directions are parallel to the spacelike separation of the end points so
that the strings compensates apart from a piece which connects the endpoints
("bridged" $q-\bar{q}$ pairs)..}

The attractive aspect of this theoretical definition of the physical meaning
of confinement is that it does not only explain the observational situation
but it can in principle also be checked in terms of an extension of the
Yennie-Frautschi-Suura infrared resummation techniques applied to the
off-shell logarithmic infrared divergencies of stringlike gluon correlations
in the massless limit $m\rightarrow0,$ using the vectormeson mass as a natural
infrared regularization parameter. In the pointlike BRST gauge setting there
are no physical massive gluon fields which one could use in such a
calculation. One expects that the Hilbert space positivity plays an essential
role in the understanding of physical aspects of infrared phenomena. It is
interesting to note that, whereas this confinement mechanism would
perturbative accessible by resummation of leading logarithmic infrared
divergencies in the $m\rightarrow0$ limit, the problem of bound states
(gluonium, hadrons as bound states of quarks) associated with pointlike
composite fields remains still outside the range of presently known
perturbative resummation methods.

An almost trivial illustration of a spectrum changing mechanism is provided by
exponentials of a massive free field $\phi(x)$ in two spacetime dimensions
whose two-pointfunction is logarithmically divergent in the massless limit
$m\rightarrow0;$ so that the perturbative expansion of correlations of the
exponential fields $exp\pm ig\phi$ lead to logarithmically infrared divergent
series in $g.$ On the other hand the exact limiting behavior is a power law.
Muliplying the exponential field with a power in the mass $m^{\alpha}%
expig\phi,$ the $a$ can be adjusted in such a way that all expectation values
of the exponential field and its Hermitian conjugate remain finite in the
massless limit. The result is the emergence of the charge conservation law:
all correlations for which the $+g$ charges do not compensate the $-g~$charges
vanish, so that only neutral correlations remain in the massless limit. The
perturbative mechanism for the infrared divergencies of the massive
gluon-quark system in the limit of vanishing gluon mass is expected to be
analogous; but since the gluons are chargeless and the $e$-stringlocal nature
has to be taken into account, the expected analogous results is that only
correlations of pointlike composites and $q-\bar{q}~$pairs with a finite
connecting string (the charge congugation inverts the $e$-direction) remain. 

If the model of QCD is really capable to describe confinement, there is no
alternative to this picture about implications of perturbative logarithmic
infrared divergencies. The necessary perturbative resummation techniques
should be similar to those used by Yenni-Frautschi-Suura \cite{YFS} to show
that the scattering amplitudes (but not the off-shell correlation functions)
of charged particicles and a finite number of photons vanish in the limit of
vanishing photon mass. In all these $m\rightarrow0~$limits it is important
that the interpolating massive theory fulfills the Hilbert space positivity
which requires the use of stringlocal fields and excludes the BRST gauge
setting whose physical range of validity does not extend beyond the gauge
invariant observables.

The stringlocal free fields of the Wigner stuff are irreducible in a very
strong sense. Whereas in the gluon/quark model the irreducibility referred to
the model-defining interacting strings whose confinement is manifested in the
vanishing of correlation functions containing gluon fields or "unbridged"
$q-\bar{q}$ pairs (see above), the absence of causality violation changes of
the Wigner stuff into normal matter explain its inert behavior. This should
again be seen in form of infrared divergencies in perturbative calculations of
interactions the stringlocal stuff.

Indeed, the attempt to use the transcendental propagators (\ref{prop}) of
fields associated with the Wigner stuff leads to severe perturbative infrared
divergencies. In this case there exists no massive model in Hilbert space
whose $m\rightarrow0$ limit describes the infinite spin representations and
could be ussed as a natural covariant infrared regularization\footnote{A
representation of the stuff as a limit of ordinary matter would require a
decreasing mass accompanied by an increasing spin; it is not known whether
such a representation is possible.}. The inert Wigner stuff and confinement
share the string-localization of their fields; but in the Wigner case it is a
property of a free field which by causality is prevented from entering an
interaction which transforms it into compact matter, whereas for QCD the
interacting gluons and quarks are confined i.e. they cannot escape except in
the form of pointlocal composites.

The free QFTs which are canonically associated with positive energy
representations of the Poincar\'{e} group fulfill both causality requirements
of the foundational causal localization principle namely the spacelike
Einstein causality and the timelike \textit{causal completeness property}
\cite{Sigma}. For models which permit a formal representation in terms of
Lagrangian quantization this is a consequence of the hyperbolic character of
the propagation of solutions of Euler-Lagrange equation, but the Wigner stuff
does not permit such a representation. In the algebraic setting of local
quantum physics \textit{causal completeness} is the equality of the
\textit{outer approximation} of an $\mathcal{O}$-localized algebra
\footnote{The algebra generated by smearing fields with spacetime
$\mathcal{O}$-supported testfunctions; $\mathcal{O}^{\prime}$ denotes the
causal complement of $\mathcal{O}^{\prime}$ and $\mathcal{O}^{\prime\prime}$
is its causal completion.} in terms of wedge-localized algebras $\mathcal{A}%
(W)$ is equal to inner approximation in terms of double-cone
(diamond)-localized algebras $\mathcal{A(D})$%
\[
\mathcal{A(O})=\mathcal{A(O}^{\prime\prime})~where~\mathcal{A(O}%
)=\cup_{\mathcal{D\subset O}}\mathcal{A(D}),~~\mathcal{A(O}^{\prime\prime
})=\cap_{W\supset\mathcal{O}}\mathcal{A}(W)
\]
For the noncompact Wigner stuff $\mathcal{O}$ is a noncompact convex region
which extends to spacelike infinity and instead of $D~$the inner approximation
is effected in terms of spacelike cones $\mathcal{C}$. 

In passing it is interesting to note that this important causal completion
property is violated in certain cases which appeared in the literature without
the protagonists of these proposals having noticed this deficiency. It is
always violated on one side in the mathematical AdS-CFT isomorphism (i.e. it
cannot hold simultaneously on both sides) and in proposals about extra
dimension and attempts to use Kaluza-Klein dimensional reductions in QFT
(outside quasiclassical approximations) \cite{Sigma} \cite{hol}.

Since causal localization is the defining principle of QFT it would be
surprising if nature misses the chance to realize its foundational principle
in the guise of Wigners continuous helicity class. As a result of its
noncompact spacetime localization and its reactive inertness apart from
gravity, its only possible arena of physical manifestation would be galaxies
and not earthly laboratories. There is presently no known theoretical physical
principle which forbids nature to manifest itself as dark (inert apart from
gravitation) matter. 

There remains the question of how such inert noncompact matter can be made
compatible with its formation in a big bang. In models in which dark matter is
identified with (known or still to be discovered) forms of ordinary (compact
localized) matter this perfecy inertness cannot be realized. Such explanations
have potential problems with astrophysical observations of upper limits which
may be too low in order to account for the gravitationally inferred dark
matter content whereas for purely gravitationally coupled matter there are no
limitations coming from the "visible" forms of matter.

\section{Gravitational coupling, concluding remarks}

As all positive energy matter, the Wigner stuff couples to the gravitational
field. The rough argument uses the effective mass obtained from the Einstein
relation $E=mc^{2}.~$It also possesses a conserved stringlocal energy-momentum
tensor (use the wave equation for $\Psi$)
\begin{equation}
T^{\mu\nu}(x,e)=~:\partial^{\mu}\Psi(x,e)\partial^{\nu}(x,e):\label{st}%
\end{equation}
whose expectation in suitable states can be used on the right hand side of the
Einstein-Hilbert field equation\footnote{The problem of how an $e$-dependent
energy-momentum tensor can be related with the generators of ($e$-independent)
spacetime transformations remains open. }. It would be interesting to consider
expectation values in quasifree states to study the induced gravitation of
noncompact matter. Note that the noncompact Wigner stuff does not permit a
pointlike energy-momentum tensor, the stringlike representation (\ref{st}) is
the tightest possible local representation.\ 

It is clear that the only possible physical use of this stuff is as a
candidate for dark matter. Unlike other dark matter candidates (WIMPS,..) it
is not an object which has been invented exclusively in order to explain dark
matter; this positive energy stuff made its debut already in Wigner's 1939
paper which was written in the same decade in which Zwicky discovered dark
matter. What was missing for more than seven decades was an understanding of
its inherent noncompact localization in terms of a field theoretic
description. \ Unlike explanations in terms of known matter there are no
astrophysical restrictions on the Wigner stuff except those coming from
galactic changes of the gravitational balance. This permits to adjust its
density to whatever it takes to obtain agreement with the measured
gravitational balance over galactic distances. In particular the observational
reasons why normal zero mass matter (photons, gravitons, massless neutrinos)
cannot account for dark matter do not apply to Wigner's stuff; its noncompact
nature provides it with the ability to "stick" to galaxies or clusters of
galaxies despite its vanishing rest mass.

As mentioned in the introduction there is no other proposal which fulfills
Poppers falsification criterion as perfect as Wigner's stuff; any
counter-registered event, for which there exist convincing reasons to believe
that it is caused by the presence of dark matter, would throw the present
proposal of identifying Wigner's stuff with galactic dark matter into the
dustbin. It seems somewhat paradoxical that it is the only kind of theoretical
positive energy matter whose verification of existence in nature depends on
its invisibility with respect to earthly particle counters. There is however
the before mentioned problem to understand how such inert noncompact stuff got
into our universe in the aftermath of a big bang.

A rough look at the observational situation (about which the author has no
astrophysical expertise) with respect to the contribution of massless matter
seems to indicate that photons, gravitons and even additional types of
massless neutrinos would not be sufficient to account for the necessary
gravitationally observed amount of dark matter. With the inclusion of the
Wigner stuff the astrophysical observational upper limit restrictions would be
eliminated since inert matter is by definition not subject to
(non-gravitational) observational restrictions; the density of the stuff can
be adjusted so that it fits the amount of gravitationally inferred dark
matter. Particle physicists who expected an explanation of dark matter in the
present work in terms of yet another kind of WIMPS/-inos will be disappointed.
The present proposal is much more fundamental but not necessarily more
acceptable. But it is the only attempt which will remain after all efforts to
identify earthly WIMPS/-inos or to place the burden on large distance
modification of Einstein/Newton gravity have failed. The Wigner stuff is a
third proposal to explain dark matter which differs significantly from the two
existing ones.  

The present proposal for dark matter is not the result of astrophysical
expertise, but rather of conceptual curiosity about Wigner's positive energy
stuff. Even if astrophysicists will be able to exclude this gravitating but
otherwise inert stuff as a contender for galactic dark matter, the historical
amazement about its theoretical discovery in the same decade as Zwicky's
observation of dark matter and the surprise about the more than 6 decades
lasting effort \cite{MSY} \cite{Pla} to unravel its possible field theoretic
physical properties will still remain. Its theoretical importance for the
understanding of the field theoretic description of all three classes of
Wigner's positive energy matter and its historic role for discovering the
relevance of string-localization of interacting ordinary matter is beyond
doubt. Until astrophysical arguments for its exclusion are found, it should be
added to the list of dark matter candidates.

\textbf{Acknowledgement}: I thank Jakob Yngvason and Jens Mund for their
continuous interest in the third Wigner class whose complete mathematical
understanding in terms of modular localization was the result of a past joint
collaboration \cite{MSY}.

I am indebted to Karl-Henning Rehren for a critical reading of the manuscript.


\begin{thebibliography}{99}                                                                                               %


\bibitem {Wig}E.P. Wigner, On unitary representations of the inhomogeneous
Lorentz group, Ann. Math. \textbf{40}, (1939)

\bibitem {Y}J. Yngvason, Zero-mass infinite spin representations of the
Poincar\'{e} group and quantum field theory, Commun. Math. Phys. \textbf{18}
(1970), 195

\bibitem {Wein}S. Weinberg, \textit{The Quantum Theory of Fields I}, Cambridge
University Press 1991

\bibitem {BGL}R. Brunetti, D. Guido and R. Longo, \textit{Modular localization
and Wigner particles}, Rev. Math. Phys. \textbf{14}, (2002) 759

\bibitem {MSY}J. Mund, B. Schroer and J. Yngvason, Commun. Math. Phys.
\textbf{268}, (2006) 621

\bibitem {stringlocal}B. Schroer, A Hilbert space setting for interacting
higher spin fields and the Higgs issue, arXiv:1407.0369

\bibitem {Hilbert}B. Schroer, \textit{A Hilbert space setting which replaces
Gauge Theory}, arXiv:1410.0782

\bibitem {Schu}P. Schuster and N. Toro, \textit{A Gauge Field Theory of
Continuous-Spin Particles}, JHEP \textbf{1310} (2013) 061, arXiv:1302.3225

\bibitem {Pla}M. Plaschke and J. Yngvason, Journal of Math. Phys. \textbf{53},
(2012) 042301

\bibitem {Haag}R. Haag, \textit{Local Quantum Physics}, Springer 1996

\bibitem {Bu-Fr}D. Buchholz and K. Fredenhagen, Nucl. Phys. B \textbf{154},
(1979) 226

\bibitem {Bu}D. Buchholz, Phys. Lett.\textbf{ B174}, (1986) 331

\bibitem {Froe}J. Fr\"{o}hlich, G. Morchio and F. Strocchi, Phys. Lett.
\textbf{89B}, (1979) 61

\bibitem {Sigma}B. Schroer, \textit{The Ongoing Impact of Modular Localization
on Particle Theory}, Sigma 10, (2014) 065

\bibitem {hol}B. Schroer, \textit{Modular localization and the holistic
structure of causal quantum theory, a historical perspective}, to be published
in SHPMP

\bibitem {YFS}D. Yenni, S. Frautschi and H. Suura, Ann. of Phys. \textbf{13},
(1961) 370
\end{thebibliography}
\end{document}